\documentclass[a4paper]{article}

\usepackage[left = 3 cm, right = 3 cm]{geometry}
\usepackage[T1]{fontenc} 
\usepackage[utf8]{inputenc}
\usepackage{lmodern}
\usepackage{bm} 
\usepackage{amsmath, amsfonts, amssymb}
\usepackage{graphicx}
\usepackage{hyperref}
\usepackage{upgreek}
\usepackage[super, sort&compress]{natbib}
\usepackage{authblk}
\usepackage{todonotes, soul, xcolor}
\usepackage{setspace}
\usepackage{multirow}
\usepackage{makecell}
\usepackage{rotating}

\author[2,*]{D\'{a}niel Tibor Pozs\'{a}r}
\author[2]{Marcell Bal\'{a}zs S\'{i}pos}
\author[4,5]{Amador Garc\'{i}a-Fuente}
\author[1]{P\'{e}ter Nemes-Incze}
\author[2]{Viktor Ivády}
\author[3]{Efren Navarro-Moratalla}
\author[4,5]{Jaime Ferrer}
\author[6,7]{L\'{a}szl\'{o} Szunyogh}
\author[2]{L\'{a}szl\'{o} Oroszl\'{a}ny}
\author[1,2,*]{Zolt\'{a}n Tajkov}

\affil[1]{Centre for Energy Research, Institute of Technical Physics and Materials Science, 1121 Budapest, Hungary}
\affil[2]{Department of Physics of Complex Systems, ELTE Eötvös Loránd University, 1117 Budapest, Hungary}
\affil[3]{Instituto de Ciencia Molecular (ICMol), Universitat de València, 46980 Spain}
\affil[4]{Departamento de Física, Universidad de Oviedo, 33007 Oviedo, Spain}
\affil[5]{Centro de Investigación en Nanomateriales y Nanotecnología, Universidad de Oviedo--Consejo Superior de Investigaciones Científicas, 33940 El Entrego, Spain}
\affil[6]{Department of Theoretical Physics, Institute of Physics, Budapest University of Technology and Economics, Műegyetem rkp.~3., HU-1111 Budapest, Hungary}
\affil[7]{HUN-REN-BME Condensed Matter Research Group, Budapest University of Technology and Economics, Műegyetem rkp.~3., HU-1111 Budapest, Hungary}

\affil[*]{\small \emph{Corresponding authors: daniel.pozsar@ttk.elte.hu, tajkov.zoltan@ek-cer.hu}}

\title{Proximity-Induced Spin Reorientation in Monolayer CrI$_3$ on Hexagonal WTe$_2$}

\begin{document}

\maketitle

\begin{abstract}
Magnetic anisotropy controls the orientation and thermal stability of two-dimensional magnetic order. Predicting proximity-induced changes in anisotropy requires linking the electronic structure to microscopic magnetic interactions and finite-temperature behavior. Here we study monolayer CrI$_3$ on hexagonal WTe$_2$ using a first-principles-to-finite-temperature workflow centered on relativistic spin-Hamiltonian mapping. We find that WTe$_2$ reorients the CrI$_3$ magnetization from out-of-plane to in-plane and substantially enhances the magnetic ordering scale within the extracted spin models. Analysis of the extracted spin Hamiltonians shows that the reorientation is driven by a substrate-induced change in the balance between symmetric anisotropic exchange and onsite anisotropy. We establish a transferable workflow for proximity-controlled magnetism in two-dimensional van der Waals heterostructures.
\end{abstract}

\doublespacing


\section{\label{sec:intro}Introduction}

\begin{figure}[t!]
    \centering
    \includegraphics[width=1\linewidth]{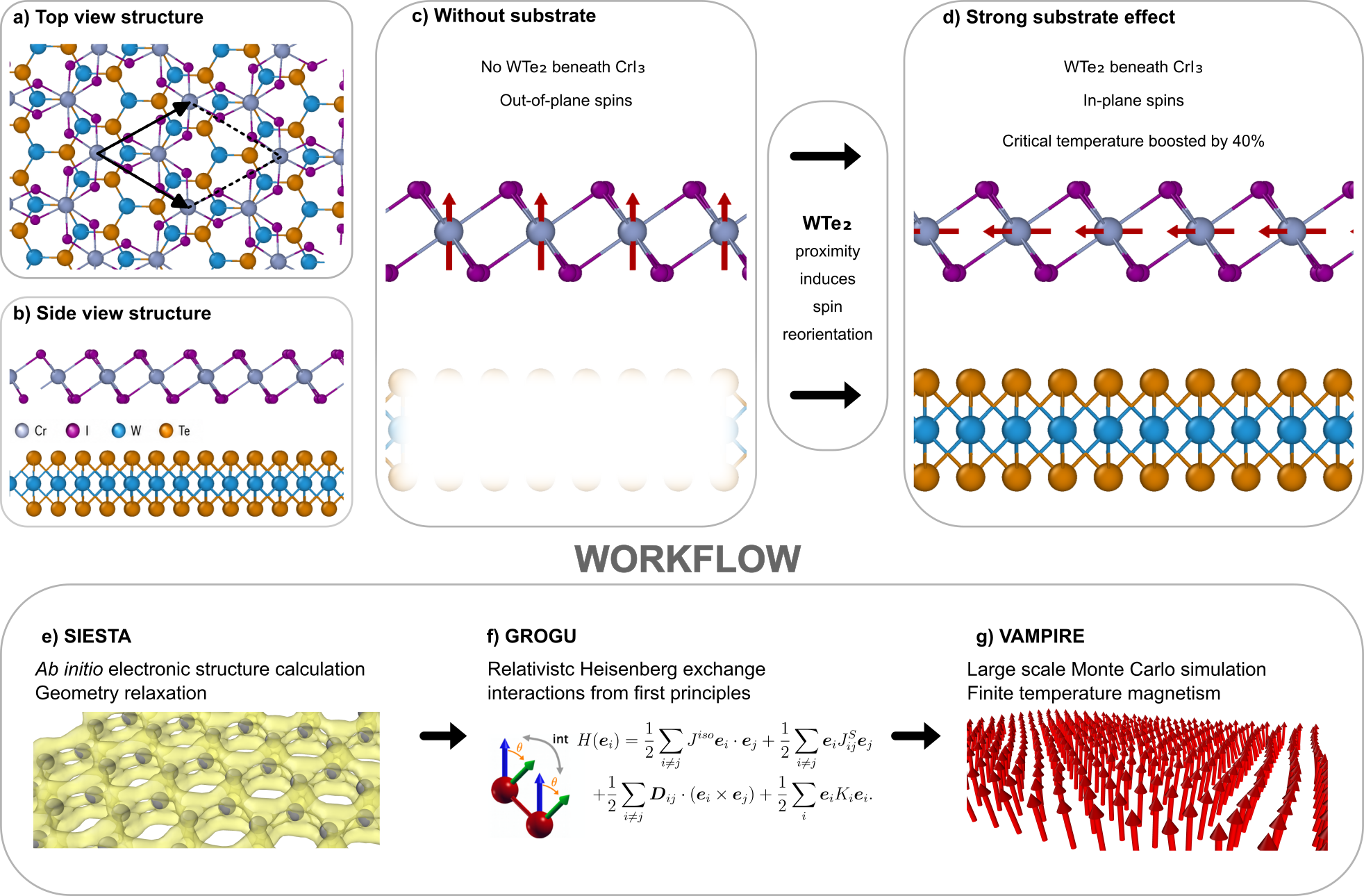}
    \caption{
\textbf{Proximity-controlled magnetism and multiscale computational workflow.}
\textbf{a,b)} Top and side views of the CrI$_3$/WTe$_2$ van der Waals heterostructure, showing the atomic registry and the two-monolayer geometry.
\textbf{c)} In the absence of the WTe$_2$ proximity effect, monolayer CrI$_3$ retains an out-of-plane magnetic orientation.
\textbf{d)} In the CrI$_3$/WTe$_2$ heterostructure, the WTe$_2$ substrate reorients the CrI$_3$ magnetization into an in-plane state.
\textbf{e--g)} The calculations follow a first-principles-to-spin-model workflow: SIESTA provides the spin-orbit-coupled electronic structure, GROGU extracts the effective relativistic spin Hamiltonian, and VAMPIRE simulates the resulting finite-temperature magnetic order.
}
    \label{fig:intro_workflow}
\end{figure}

The discovery of intrinsic magnetism in atomically thin van der Waals crystals has established two-dimensional magnets as a platform for exploring magnetic order in the reduced-dimensional limit and for building spintronic heterostructures with layer-by-layer control \cite{Gong2017_CrGeTe3Intrinsic2DFerromagnetism,Huang2017_LayerDependentCrI3Ferromagnetism,Burch2018_Magnetism2DvdWMaterials,Gibertini2019_Magnetic2DMaterialsHeterostructures}. In this setting, magnetic anisotropy is not a secondary material parameter: it selects the easy direction of the ordered moments, opens the spin-wave gap needed to stabilize long-range order in two dimensions, and strongly influences the finite-temperature ordering scale\cite{Burch2018_Magnetism2DvdWMaterials,Gibertini2019_Magnetic2DMaterialsHeterostructures,Lado2017_CrI3MagneticAnisotropyOrigin,Tiwari2021_CriticalBehaviorCrI3CrBr3CrGeTe3FeCl2,Pavizhakumari2025_BeyondRPACurieTemperatureCrI3}. Monolayer CrI$_3$ has become a paradigmatic example, because it combines a well-defined out-of-plane ferromagnetic ground state with a magnetic anisotropy that is sensitive to spin-orbit coupling, lattice geometry, and the local electronic environment \cite{Huang2017_LayerDependentCrI3Ferromagnetism,Lado2017_CrI3MagneticAnisotropyOrigin,Kashin2020_OrbitallyResolvedCrI3Ferromagnetism,JaeschkeUbiergo2021_CrI3MagnetismTheory}.
The underlying ferromagnetic exchange scale is also sensitive to the
treatment of electronic correlations. Recent DFT+U calculations including
spin--orbit coupling found a nonmonotonic dependence of the dominant
ferromagnetic exchange interaction on \(U\), identifying an
intermediate-correlation regime around \(U=2\)--\(3\) eV
\cite{Krindges2026_ElectronicCorrelationsCrI3}.

In CrI$_3$ and related chromium trihalide monolayers, this sensitivity has motivated several routes for tuning the magnetic ground state, including strain, lattice deformation, electric fields, and stacking registry \cite{WebsterYan2018_CrX3StrainMAE,Vishkayi2020_CrI3StrainElectricField,Wu2019_CrI3Strain,Pizzochero2020_CrI3LatticeDeformations,WangSanyal2021_CrI3Stacking}. Van der Waals heterostructures provide a complementary control knob: by placing a two-dimensional magnet next to a nonmagnetic layer with strong spin-orbit coupling, one can modify the relativistic magnetic interactions through interfacial hybridization and spin-orbit proximity, without necessarily imposing strong covalent bonding or large structural distortion \cite{Dolui2020_ProximitySpinOrbitTorqueCrI3,Zollner2019_ProximityExchangeMoSe2WSe2CrI3,Zollner2023_ValleySplittingTwistingGatingCrI3}. Heavy transition-metal dichalcogenides such as WTe$_2$ are particularly attractive in this context, because they introduce strong spin-orbit coupling into van der Waals interfaces \cite{Zhao2020_MagneticProximityWTe2HelicalEdge,Staros2024_WTe2CrI3TopologicalSpinFilter}. Modern deterministic transfer techniques now allow WTe$_2$ to be integrated with layered ferromagnets, enabling magnetic-proximity and spin--orbit-torque phenomena in all-van der Waals heterostructures, including efficient field-free magnetization switching \cite{Zhou2026_WTe2Fe3GaTe2SOT,Zhang2026_vdWSOTImaging,Ning2024_WTe2Fe3GeTe2SpinTransfer}. Beyond torque generation, WTe$_2$ has also been shown to enhance the magnetic ordering temperature of adjacent two-dimensional magnets through proximity-induced lattice reconstruction and interfacial charge transfer \cite{Herling2026_CGTWTe2Ferromagnetism}. The resulting changes in magnetic criticality and anisotropy can be monitored electrically through magnetotransport measurements \cite{Li2022_ProximityMagnetizedQSHI,HenriquezGuerra2025_CrSBrStrainMRMAE}, as well as optically through Kerr or magnetic circular dichroism measurements \cite{Cheng2022_ElectricFieldTunableInterfacialFerromagnetism,Eom2023_VoltageControlMagnetism}.

A central challenge is that proximity-induced changes in magnetic anisotropy are often comparable to the small relativistic energy scales that control spin-orbit-coupled magnetic interactions in CrI$_3$ and related chromium trihalide monolayers \cite{Lado2017_CrI3MagneticAnisotropyOrigin,Bacaksiz2021_CrI3CrBr3SOCMagneticProperties,Sabani2025_BeyondOrbitallyResolvedExchangeCrI3NiI2}. Total-energy comparisons between different magnetic orientations can identify an easy axis, but they do not by themselves reveal which microscopic magnetic interactions are modified, nor do they determine whether the effect survives thermal fluctuations. The missing link is a relativistic spin-Hamiltonian mapping that converts the electronic structure into isotropic exchange, symmetric anisotropic exchange, antisymmetric exchange, and onsite anisotropy terms. We perform this step with GROGU\cite{zerothi_grogupy}, a post-processing framework that extracts these interaction tensors from nonorthogonal pseudoatomic-orbital Hamiltonians using an LKAG-based torque formalism \cite{Liechtenstein1987_LKAGExchange,MartinezCarracedo2023_RelativisticMagneticInteractions,zerothi_grogupy}. The resulting spin Hamiltonians can then be used directly in finite-temperature atomistic simulations, allowing proximity-induced anisotropy changes to be followed from the electronic-structure level to the magnetic ordering scale.

Here we apply this strategy to monolayer CrI$_3$ on hexagonal WTe$_2$, as summarized in Fig.~\ref{fig:intro_workflow}. We focus on the hexagonal polymorph as a low-strain model heavy-SOC substrate, rather than as the thermodynamic ground-state phase of bulk WTe$_2$. Although WTe$_2$ is most commonly associated with the distorted 1T$^\prime$/T$_d$ structure, hexagonal monolayer WTe$_2$ has been discussed theoretically as a metastable transition-metal dichalcogenide phase and recent work has reported selective fabrication of monolayer 1H- and 1T$^\prime$-WTe$_2$ \cite{Ma2016_HexagonalTMDQuantumSpinHall,Ando2024_SelectiveFabricationMonolayerWTe2}. This choice allows us to isolate the effect of a nearly lattice-matched heavy-element van der Waals substrate on the CrI$_3$ magnetic interactions.

Using our electronic-structure--to--spin-model workflow, with GROGU providing the central spin-Hamiltonian mapping step, we find that WTe$_2$ reorients the CrI$_3$ magnetization into an in-plane state and enhances the finite-temperature ordering scale by approximately $40$--$50\%$ within the extracted spin models.

\section{Results}

\subsection{The electronic structure}

We first constructed a commensurate CrI$_3$/WTe$_2$ heterostructure from separately relaxed monolayer CrI$_3$ and hexagonal WTe$_2$ reference structures. In our DFT relaxations, the two lattices match particularly well in the hexagonal setting: the relaxed CrI$_3$ lattice constant and twice the primitive h-WTe$_2$ lattice constant differ by only 0.14\%. This small mismatch is important because magnetic exchange interactions, magnetic ground states, and magnetocrystalline anisotropy in CrI$_3$ and related chromium trihalide monolayers are known to be highly sensitive to lattice deformation and strain \cite{WebsterYan2018_CrX3StrainMAE,Vishkayi2020_CrI3StrainElectricField,Wu2019_CrI3Strain,Pizzochero2020_CrI3LatticeDeformations}. The resulting heterostructure was represented by a 20-atom hexagonal simulation cell containing one CrI$_3$ layer with 2 Cr and 6 I atoms and one h-WTe$_2$ layer with 4 W and 8 Te atoms. The common in-plane lattice constant was $a=7.1303~\AA{}$, with a $30~\AA{}$ out-of-plane cell height used to separate periodic images. Since magnetic couplings in layered CrI$_3$ systems can also depend sensitively on stacking registry \cite{WangSanyal2021_CrI3Stacking}, the relative in-plane displacement and interlayer separation were selected from lateral and vertical stacking scans. The final structural model and the stacking-energy landscape are reported in Supplementary Section 1.

Having established the structural model, we next examined how the WTe$_2$ substrate modifies the preferred magnetic orientation of the CrI$_3$ layer. In all magnetic configurations considered here, the two Cr moments remain nearly ferromagnetically aligned, while the WTe$_2$ layer carries only a small induced spin polarization. We therefore characterize the magnetic anisotropy by comparing self-consistent spin-orbit DFT calculations initialized with different global magnetization directions: an out-of-plane configuration (OOP) and a set of in-plane configurations (IP) with different azimuthal angles. The detailed orientation scan, including the final spin directions and total energies, is reported in Supplementary Section 2.

Figure~\ref{fig:dft_anisotropy_bands}a summarizes the central result of this scan in terms of
$\Delta E = E_{\rm IP}-E_{\rm OOP}$, where positive values indicate an out-of-plane easy orientation and negative values indicate an in-plane one. For the isolated CrI$_3$ control, the in-plane configurations lie $0.94~\mathrm {meV}$ per cell above the out-of-plane state, confirming the expected out-of-plane easy axis. In contrast, in the CrI$_3$/WTe$_2$ heterostructure the lowest in-plane configuration lies approximately $0.35~\mathrm{meV}$ per cell below the out-of-plane state. Thus, the WTe$_2$ substrate reverses the sign of the magnetic anisotropy, converting the CrI$_3$ layer from an out-of-plane Ising-like ferromagnet into an in-plane anisotropic ferromagnet.

\begin{figure}[t!]
    \centering
    \includegraphics[width=1\linewidth]{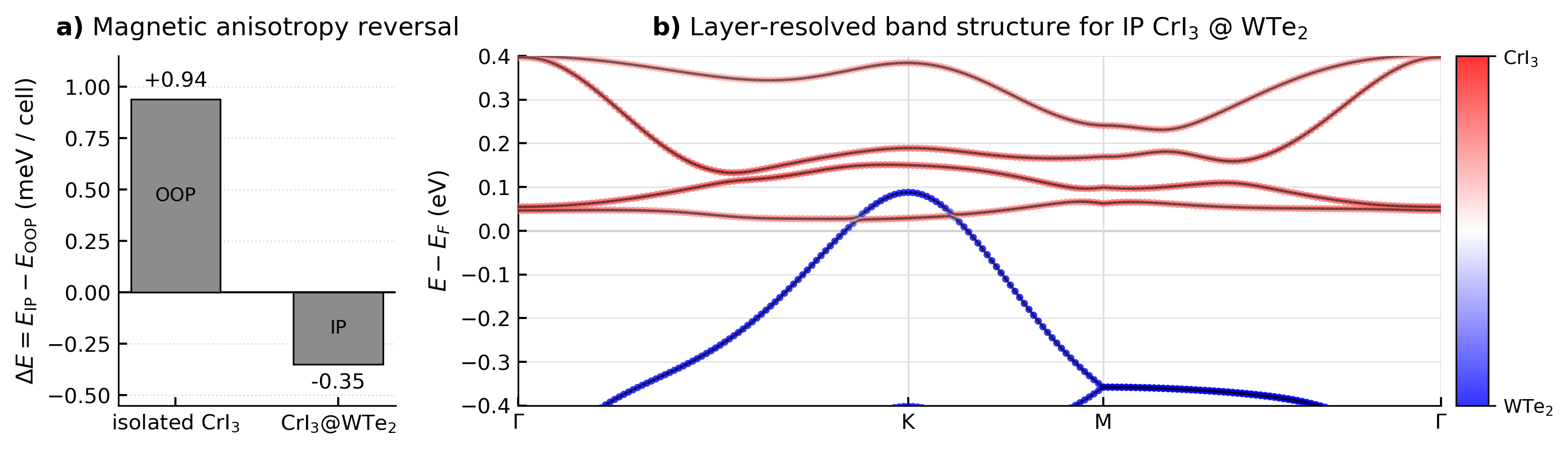}
    \caption{
\textbf{DFT-level magnetic anisotropy reversal and layer-resolved band character.}
\textbf{a)} Spin-orbit DFT energy difference $\Delta E = E_{\rm IP} - E_{\rm OOP}$ for isolated CrI$_3$ and CrI$_3$/WTe$_2$. 
The energy difference is defined separately for each system by comparing self-consistent calculations with different magnetic orientations for the same structural model; the bars therefore indicate the preferred spin orientation within each system. Positive and negative values correspond to out-of-plane and in-plane easy orientations, respectively. 
\textbf{b)} Layer-resolved band structure of the in-plane CrI$_3$/WTe$_2$ heterostructure. 
The color scale encodes the layer character, from WTe$_2$-dominated states in blue to CrI$_3$-dominated states in red.
}
    \label{fig:dft_anisotropy_bands}
\end{figure}

The layer-resolved band structure in Fig.~\ref{fig:dft_anisotropy_bands}b provides the corresponding electronic-structure context for the in-plane heterostructure. The low-energy bands retain a clear layer character: dispersive states close to the Fermi level are predominantly associated with the WTe$_2$ layer, whereas flatter CrI$_3$-derived bands remain identifiable in the same energy window. At the same time, the mixing of the layer projections near the Fermi level indicates weak interlayer hybridization, consistent with a proximity effect rather than a completely decoupled bilayer. A detailed comparison of the out-of-plane and in-plane projected band structures and PDOS is given in Supplementary Section 3. These data show only minor changes upon rotating the magnetic orientation, supporting the interpretation that the spin reorientation is not driven by a major electronic reconstruction, but by a subtle spin-orbit-induced modification of the magnetic anisotropy.

The spin reorientation does not involve a collapse of the local Cr moments. In the Mulliken spin populations, the two Cr atoms carry a combined moment of about $7.65~\mu_{\rm B}$ in the isolated CrI$_3$ reference, corresponding to approximately $3.8~\mu_{\rm B}$ per Cr atom. In the CrI$_3$/WTe$_2$ heterostructure this value remains essentially unchanged, with a combined Cr moment of about $7.71~\mu_{\rm B}$ for both the out-of-plane and in-plane magnetic configurations. The iodine ligands carry an antiparallel spin polarization of about $-0.26~\mu_{\rm B}$ per I atom, consistent with the covalent Cr--I magnetic environment of CrI$_3$. The WTe$_2$ layer remains only weakly spin-polarized, with an induced moment of order $10^{-2}~\mu_{\rm B}$ per cell. Thus, the substrate rotates the preferred orientation of a robust CrI$_3$ ferromagnetic moment rather than quenching the Cr moments or transferring substantial magnetization to the WTe$_2$ layer; the detailed Mulliken charge and spin-population tables are provided in Supplementary Section 4.

\subsection{Relativistic spin-Hamiltonian mapping}

The DFT calculations establish the substrate-induced change of the preferred magnetic orientation, but a finite-temperature description requires a microscopic spin model. We therefore used GROGU\cite{zerothi_grogupy} to map the electronic structure onto a relativistic Heisenberg Hamiltonian containing isotropic exchange, symmetric anisotropic exchange, antisymmetric exchange, and onsite anisotropy terms. This step provides the connection between the electronic-structure origin of the anisotropy and the atomistic Monte Carlo simulations discussed below.

In the convention used here, the magnetic energy is written in terms of normalized spin-direction vectors ($\mathbf e_i$) as
$$
\mathcal H =
\sum_i \mathbf e_i^{\mathsf T}\mathbf K_i \mathbf e_i
+
\frac{1}{2}\sum_{i\ne j}
\mathbf e_i^{\mathsf T}\mathbf J_{ij}\mathbf e_j ,
$$
where $\mathbf K_i$ is the onsite anisotropy tensor and $\mathbf J_{ij}$ is the full relativistic pair-exchange tensor. The latter can be decomposed into an isotropic part ($J_{ij}^{\rm iso}$), a symmetric traceless tensor $(\mathbf J_{ij}^{\rm S}$), and an antisymmetric component represented by the Dzyaloshinskii--Moriya vector ($\mathbf D_{ij}$). This decomposition allows us to separate the exchange scale controlling magnetic ordering from the relativistic anisotropic terms that determine the preferred spin orientation.

The exchange and anisotropy tensors were obtained with GROGU using the Liechtenstein--Katsnelson--Antropov--Gubanov torque formalism \cite{Liechtenstein1987_LKAGExchange}, in its relativistic extension \cite{Udvardi2003_RelativisticSpinWaves} to the nonorthogonal pseudoatomic-orbital basis \cite{MartinezCarracedo2023_RelativisticMagneticInteractions}. In this approach, local infinitesimal rotations of the exchange field are applied to the Kohn--Sham Hamiltonian, and the resulting second-order energy variations, evaluated from the Green's function, are matched to the corresponding variations of the relativistic spin Hamiltonian to extract the full pair-exchange tensors $\mathbf J_{ij}$ and onsite anisotropy tensors $\mathbf K_i$.

\begin{figure}[t!]
    \centering
    \includegraphics[width=0.75\linewidth]{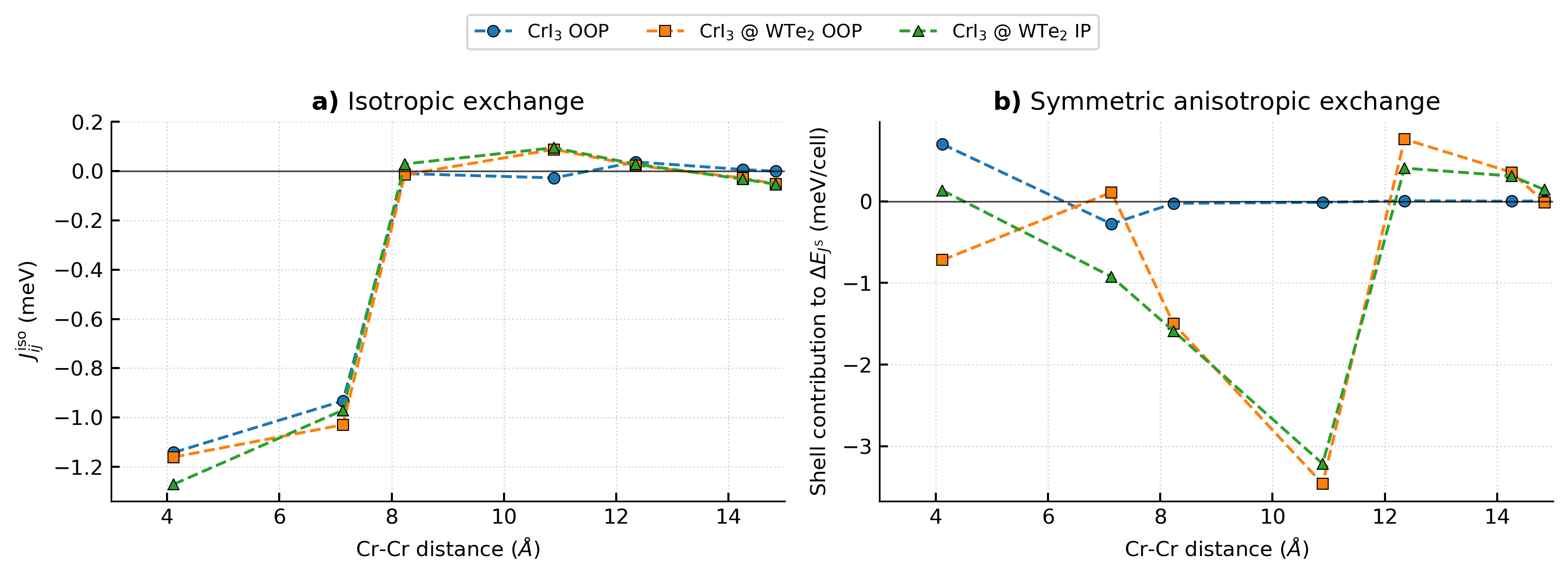}
    \caption{
\textbf{Distance dependence of isotropic and anisotropic exchange interactions.}
\textbf{a)} Shell-averaged isotropic exchange interactions, $J_{ij}^{\mathrm{iso}}$, extracted with GROGU as a function of Cr--Cr distance for isolated CrI$_3$, CrI$_3$/WTe$_2$ with an out-of-plane reference configuration, and CrI$_3$/WTe$_2$ with an in-plane reference configuration.
Negative values favor ferromagnetic alignment in the Hamiltonian convention used here.
The dominant ferromagnetic nearest-neighbour exchange couplings are similar in the three parameter sets, while the WTe$_2$ substrate modifies weaker longer-range couplings.
\textbf{b)} Shell-resolved contribution of the symmetric traceless exchange tensor to the relativistic spin-Hamiltonian orientation-energy difference, $\Delta E_{J^{\rm S}}=E_{J^{\rm S}}(\mathbf e_{\rm IP})-E_{J^{\rm S}}(\mathbf e_{\rm OOP})$, evaluated for $\mathbf e_{\rm OOP}=(0,0,1)$ and $\mathbf e_{\rm IP}=(\cos60^\circ,\sin60^\circ,0)$.
Negative values favor the in-plane orientation.
The heterostructure develops negative contributions over several Cr--Cr shells, showing that the substrate-induced reorientation is encoded directly in the anisotropic two-ion exchange rather than arising from an onsite anisotropy flip alone.
}
    \label{fig:jiso_GROGU}
\end{figure}

We first analyze the isotropic part of the pair-exchange tensor, $J_{ij}^{\rm iso}=\mathrm{Tr}\,\mathbf J_{ij}/3$, because it sets the dominant exchange scale entering the finite-temperature spin simulations. With the Hamiltonian convention used above, negative $J_{ij}^{\rm iso}$ values favor ferromagnetic alignment. Figure~\ref{fig:jiso_GROGU}a shows the shell-averaged $J_{ij}^{\rm iso}$ as a function of Cr--Cr distance for isolated CrI$_3$ and for the CrI$_3$/WTe$_2$ heterostructure, using both out-of-plane and in-plane reference configurations for the latter. In all cases, the leading exchange interactions remain ferromagnetic, indicating that the WTe$_2$ substrate does not destroy the underlying ferromagnetic CrI$_3$ exchange network. At the same time, the substrate modifies the longer-range part of the exchange profile. In isolated CrI$_3$, the shells beyond the leading ferromagnetic interactions are mostly small, with $J_{ij}^{\rm iso}$ values close to zero on the scale of the dominant couplings. In the heterostructure, several more distant shells acquire appreciable values: for example, the shells near $14.3$ and $14.8$~\AA{} become ferromagnetic, while the shell near $10.9$~\AA{} changes in the opposite direction. Thus, the WTe$_2$ layer does not simply rescale the nearest-neighbour exchange, but redistributes the isotropic exchange over a longer range. This modified exchange profile contributes to the enhanced ordering scale found in the Monte Carlo simulations. The origin of the substrate-induced spin reorientation is instead revealed by the anisotropic exchange decomposition discussed next.

We next evaluated how the relativistic terms of the extracted spin Hamiltonian encode the preferred magnetic orientation. For each parameter set, we compared two uniform ferromagnetic configurations: an out-of-plane state, $\mathbf e_{\rm OOP}=(0,0,1)$, and an in-plane state, $\mathbf e_{\rm IP}=(\cos\phi,\sin\phi,0)$, with $\phi=60^\circ$. We define the relativistic spin-Hamiltonian orientation-energy difference as
\[
\Delta E_{\rm SH}^{\rm rel}
=
E_{\rm SH}^{\rm rel}(\mathbf e_{\rm IP})
-
E_{\rm SH}^{\rm rel}(\mathbf e_{\rm OOP}),
\]
so that positive values indicate an out-of-plane preference and negative values indicate an in-plane preference. This quantity is evaluated within the extracted spin Hamiltonian and should be distinguished from the self-consistent DFT magnetic anisotropy energy discussed above. We decompose it into the symmetric traceless pair-exchange contribution, $\Delta E_{J^{\rm S}}$, and the onsite anisotropy contribution, $\Delta E_K$:
\[
\Delta E_{\rm SH}^{\rm rel}
=
\Delta E_{J^{\rm S}}
+
\Delta E_K .
\]
The symmetric traceless exchange contribution is evaluated as
\[
\Delta E_{J^{\rm S}}
=
\frac{1}{2}
\sum_{i\ne j}
\left[
\mathbf e_{\rm IP}^{\mathsf T}
\mathbf J^{\rm S}_{ij}
\mathbf e_{\rm IP}
-
\mathbf e_{\rm OOP}^{\mathsf T}
\mathbf J^{\rm S}_{ij}
\mathbf e_{\rm OOP}
\right],
\]
where the factor of $1/2$ avoids double counting of directed pair entries. The onsite contribution is evaluated as
\[
\Delta E_K
=
\sum_i
\left[
\mathbf e_{\rm IP}^{\mathsf T}
\mathbf K_i
\mathbf e_{\rm IP}
-
\mathbf e_{\rm OOP}^{\mathsf T}
\mathbf K_i
\mathbf e_{\rm OOP}
\right].
\]

The shell-resolved contributions to $\Delta E_{J^{\rm S}}$ are shown in Fig.~\ref{fig:jiso_GROGU}b. In isolated CrI$_3$, the symmetric anisotropic exchange gives a small positive net contribution. In the heterostructure, by contrast, several Cr--Cr shells contribute negatively, producing the large in-plane-favoring $\Delta E_{J^{\rm S}}$ reported in Table~\ref{tab:GROGU_anisotropy}.

\begin{table}[t]
\centering
\caption{
\textbf{Relativistic orientation-energy contributions from the extracted spin Hamiltonians.}
The listed values are orientation-energy differences
$\Delta E = E_{\rm IP}-E_{\rm OOP}$ evaluated within each extracted
spin-Hamiltonian parameter set, using
$\mathbf e_{\rm OOP}=(0,0,1)$ and
$\mathbf e_{\rm IP}=(\cos\phi,\sin\phi,0)$ with $\phi=60^\circ$.
Positive values indicate an out-of-plane preference, whereas negative values
indicate an in-plane preference. The entries decompose the anisotropic part of
the Hamiltonian into the symmetric traceless exchange contribution,
$\Delta E_{J^{\rm S}}$, and the onsite anisotropy contribution, $\Delta E_K$.
These values are not absolute DFT total-energy differences between different
systems, but orientation energies evaluated for each individual spin-Hamiltonian
model. Full bond-resolved tensors and onsite anisotropy matrices are provided in the accompanying data repository.
}
\small
\begin{tabular}{lccc}
\hline
 & CrI$_3$ & CrI$_3$/WTe$_2$ & CrI$_3$/WTe$_2$ \\
 & (OOP ref.) & (OOP ref.) & (IP ref.) \\
\hline
$\Delta E_{J^{\rm S}}$ (meV/cell) & $+0.397$ & $-4.360$ & $-3.751$ \\
$\Delta E_K$ (meV/cell)           & $+0.158$ & $+0.859$ & $+0.851$ \\
$\Delta E_{\rm SH}^{\rm rel}$ (meV/cell) & $+0.555$ & $-3.501$ & $-2.899$ \\
Preferred orientation             & OOP      & IP       & IP \\
\hline
\end{tabular}
\label{tab:GROGU_anisotropy}
\end{table}

Table~\ref{tab:GROGU_anisotropy} summarizes the corresponding integrated anisotropy contributions. The isolated CrI$_3$ reference has a positive anisotropic orientation energy, with both the symmetric traceless exchange and onsite anisotropy terms favoring the out-of-plane direction. In the CrI$_3$/WTe$_2$ heterostructure, however, the onsite anisotropy contribution remains positive and even increases to about $0.85$~meV/cell, meaning that this term alone would still favor an out-of-plane orientation. The sign reversal instead comes from the symmetric traceless exchange contribution, which becomes strongly negative and overcomes the onsite anisotropy for both heterostructure reference states. Thus, within the extracted spin Hamiltonians, the proximity-induced reorientation is not a simple flip of the onsite anisotropy, but is primarily encoded in the relativistic two-ion anisotropy.

The antisymmetric part of the exchange tensor also becomes finite in the heterostructure, as expected for a spin-orbit-coupled interface with broken inversion symmetry. However, the Dzyaloshinskii--Moriya term does not directly contribute to the energy difference between strictly collinear ferromagnetic out-of-plane and in-plane configurations, because $\mathbf e_i \times \mathbf e_j=\mathbf 0$ for a uniform ferromagnet. The substrate-induced spin reorientation discussed here is therefore traced primarily to the competition between the symmetric traceless exchange and onsite anisotropy terms, rather than to a DMI-driven noncollinear mechanism. The full bond-resolved exchange tensors and onsite anisotropy matrices are provided in the accompanying data repository.

\subsection{Finite-temperature magnetic order}

For each parameter set, the Cr magnetic moments were represented by classical unit spins placed at the Cr positions of the structural model and periodically replicated in the simulation cell. The simulations retained the bond-resolved exchange tensors extracted with GROGU, together with the onsite anisotropy term. Since the heterostructure favors an in-plane orientation, magnetic order was characterized using the total normalized magnetization, $m=|\mathbf M|/M_0$, rather than only its out-of-plane component. The ordering scale was estimated from the peak of the corresponding magnetic susceptibility; simulation-size and protocol details are reported in Supplementary Section 6.

\begin{figure}[t!]
    \centering
    \includegraphics[width=1\linewidth]{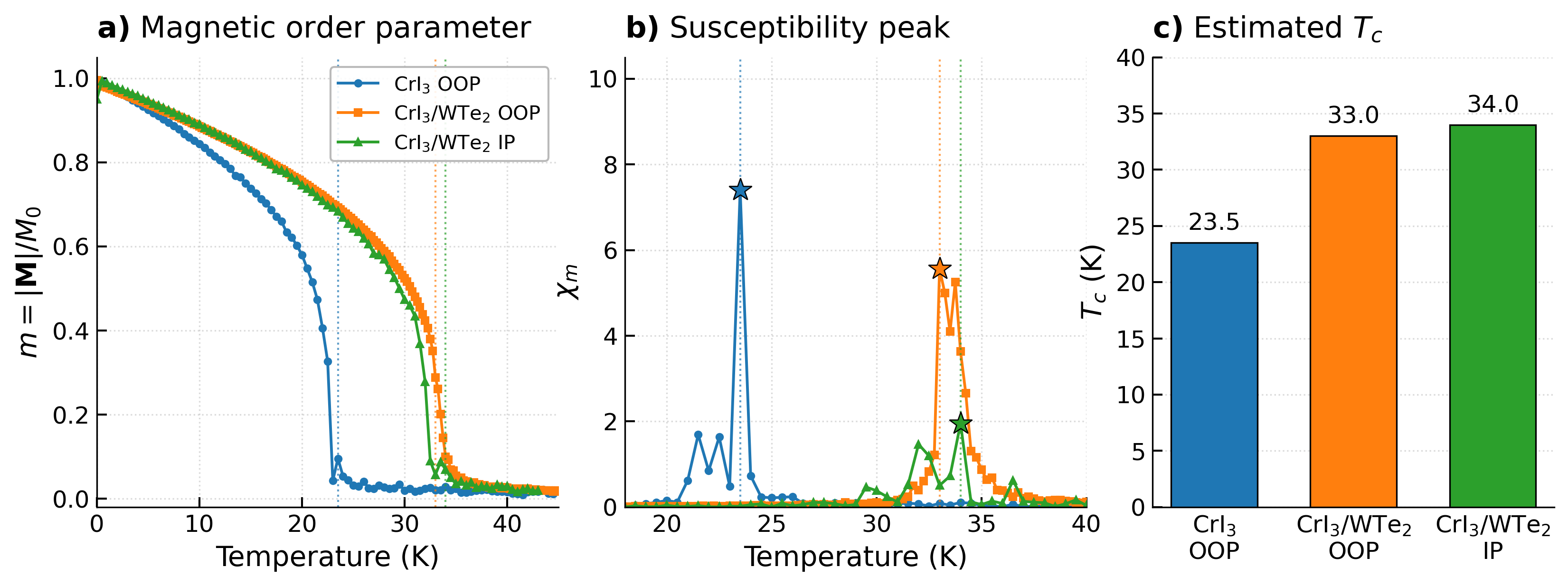}
    \caption{
\textbf{Finite-temperature magnetic order from VAMPIRE simulations.}
\textbf{a)} Normalized magnetic order parameter,
$m=|\mathbf{M}|/M_0$, as a function of temperature for isolated CrI$_3$ and
CrI$_3$/WTe$_2$ spin-Hamiltonian models.
\textbf{b)} Magnetic susceptibility $\chi_m$ near the transition region. Stars
and vertical dotted lines mark the susceptibility maxima used to estimate the
ordering temperatures.
\textbf{c)} Extracted ordering temperatures. The isolated CrI$_3$ model gives
$T_c=23.5$ K, while the CrI$_3$/WTe$_2$ models give higher transition
temperatures of $33.0$ K and $34.0$ K for the OOP- and IP-reference
Hamiltonians, respectively.
}
    \label{fig:vampire_tc}
\end{figure}

Figure~\ref{fig:vampire_tc} shows the resulting temperature-dependent magnetization and susceptibility curves. For the isolated CrI$_3$ reference model, the susceptibility peak gives an ordering temperature of approximately $T_c=23.5$~K, consistent with the weak but finite anisotropy required to stabilize two-dimensional magnetic order. In contrast, the CrI$_3$/WTe$_2$ parameter sets yield substantially higher ordering temperatures: $T_c\approx 33.0$~K for the heterostructure Hamiltonian extracted from the out-of-plane reference state and $T_c\simeq 34.0$~K for the one extracted from the in-plane reference state. Thus, the WTe$_2$ substrate not only reorients the preferred magnetization direction of the CrI$_3$ layer, but also increases the finite-temperature magnetic ordering scale by roughly $40$--$50\%$ within the extracted spin models.

The two heterostructure parameter sets give very similar ordering scales despite being extracted from different magnetic reference configurations, which supports the robustness of the substrate-induced enhancement. Moreover, the ordered phase obtained in the heterostructure simulations retains the in-plane character encoded in the spin Hamiltonian, showing that the anisotropy reversal is not washed out by thermal fluctuations near the transition. We therefore interpret the Monte Carlo results as the finite-temperature manifestation of the same proximity mechanism identified at the DFT and spin-Hamiltonian levels: WTe$_2$ modifies the relativistic anisotropic interactions of CrI$_3$, stabilizing an in-plane magnetic state while increasing the magnetic ordering scale.

We emphasize that the absolute ordering temperatures obtained here should be interpreted as ordering scales of the extracted classical spin models, rather than as direct quantitative predictions of the experimental monolayer CrI$_3$ Curie temperature. Similar first-principles exchange-mapping studies combined with classical Monte Carlo simulations have reported monolayer CrI$_3$ ordering temperatures in the same range, for example $T_c=28.0$~K in Ref.~\cite{WangSanyal2021_CrI3Stacking}, below the experimental value of about $45$~K. In the present work, the central comparison is therefore the relative change obtained within one consistent protocol: the WTe$_2$ substrate increases the ordering scale by approximately $40$--$50\%$ while simultaneously changing the preferred magnetic orientation.

Additional finite-size checks, susceptibility curves, and simulation protocol details are provided in Supplementary Section 6.

\section*{Discussion}

This work demonstrates a first-principles-to-finite-temperature workflow for proximity-controlled magnetism in a two-dimensional van der Waals heterostructure. Starting from a low-strain CrI$_3$/WTe$_2$ structural model, spin-orbit-coupled DFT calculations identify the substrate-induced change in magnetic orientation, GROGU maps the electronic structure onto a relativistic spin Hamiltonian, and  atomistic Monte Carlo simulations propagate the resulting interactions to finite temperature. 

For the CrI$_3$/WTe$_2$ system studied here, the WTe$_2$ substrate converts the out-of-plane easy orientation of isolated CrI$_3$ into an in-plane magnetic state while increasing the ordering scale obtained from the extracted spin models. The electronic-structure analysis shows that the Cr moments remain robust and that the low-energy bands are modified mainly through weak interlayer hybridization. The GROGU decomposition then identifies the key microscopic change as a substrate-induced modification of the relativistic anisotropic interactions, in particular the symmetric traceless exchange contribution, which overcomes the onsite anisotropy that would otherwise favor the out-of-plane direction. 

More broadly, these results highlight the importance of treating spin-orbit proximity effects beyond total-energy comparisons alone. In two-dimensional magnets, where magnetic anisotropies and ordering temperatures are controlled by small relativistic energy scales, a predictive description requires connecting the electronic-structure origin of anisotropy to finite-temperature magnetic behavior. The  workflow used here provides such a route and can be applied to other van der Waals magnetic heterostructures in which heavy-element substrates, interfaces, or stacking control the balance between magnetic orientation and thermal stability.

\section*{Methods}

\subsection*{SIESTA}

All first-principles calculations were performed with the SIESTA code using norm-conserving pseudopotentials and numerical pseudoatomic-orbital basis sets \cite{Soler2002_SIESTA,Garcia2020_SIESTA}. Exchange and correlation were treated within the generalized-gradient approximation using the Perdew--Burke--Ernzerhof functional \cite{Perdew1996_PBE}. Spin-orbit coupling was included self-consistently in the noncollinear formalism with the full spin-orbit strength.

 Brillouin-zone integrations for the spin-orbit DFT calculations were performed using a $30\times30\times1$ Monkhorst--Pack $k$-point grid. A fine real-space integration grid corresponding to a mesh cutoff of $4000$~Ry was used throughout the production calculations.

The numerical atomic-orbital basis was constructed using an energy shift of $10$~meV, with PAO split norms of $0.15$. The basis included double-$\zeta$ Cr $4s$ and $3d$ orbitals, single-$\zeta$ Cr $4p$ orbitals, double-$\zeta$ I $5s$ and $5p$ orbitals with polarization functions, triple-$\zeta$ W $6s$ and $5d$ orbitals together with double-$\zeta$ W $6p$ orbitals, and triple-$\zeta$ Te $5s$ and $5p$ orbitals with polarization functions. Electronic minimization was performed by direct diagonalization with an electronic temperature of $0.9$~K. The self-consistent-field cycle used Hamiltonian mixing with Pulay acceleration, a mixing weight of $0.015$, and convergence thresholds of $5\times10^{-6}$ for the density matrix and $8\times10^{-5}$~eV for the Hamiltonian.

The Cr 3d states were treated with a DFT+U correction using
\(U=3.0\) eV and \(J=0\) eV. This choice is consistent with recent
DFT+U calculations including spin--orbit coupling for monolayer
CrI$_3$, which found a nonmonotonic dependence of the dominant
ferromagnetic exchange interaction on \(U\) and identified
\(U\simeq2\)--\(3\) eV as the relevant intermediate-correlation regime
\cite{Krindges2026_ElectronicCorrelationsCrI3}.

Structural models were obtained from the relaxed monolayer geometries and the stacking/interlayer-distance search described in the Supplementary Information. The final spin-orbit calculations reported in the main text were performed for the selected structural model with fixed lattice vectors and fixed atomic coordinates. The conjugate-gradient relaxation settings used in the structural workflow employed a maximum force tolerance of $0.009$~eV/\AA{} and fixed the out-of-plane cell vector and cell angle. Mulliken charge and spin populations, orbital moments, forces, coordinates, Hamiltonian, overlap, and density-matrix files were written for post-processing.

All input files and structural files needed to reproduce the SIESTA calculations are provided in the Supplementary Information and in the accompanying data repository.

\subsection*{GROGU}

Relativistic spin-Hamiltonian parameters were extracted from the converged SIESTA spin-orbit calculations using GROGU. The mapping follows the Liechtenstein--Katsnelson--Antropov--Gubanov torque formalism \cite{Liechtenstein1987_LKAGExchange}, generalized to nonorthogonal pseudoatomic-orbital basis sets \cite{Udvardi2003_RelativisticSpinWaves} and relativistic Hamiltonians \cite{MartinezCarracedo2023_RelativisticMagneticInteractions}. Starting from the SIESTA Hamiltonian, overlap, and density-matrix files, GROGU evaluates the response of the Kohn--Sham system to local infinitesimal rotations of the magnetic exchange field and matches the resulting second-order energy variations to a generalized relativistic Heisenberg Hamiltonian.

Three spin-Hamiltonian parameter sets were generated: an isolated CrI$_3$ reference with out-of-plane magnetization, a CrI$_3$/WTe$_2$ heterostructure reference initialized out of plane, and a CrI$_3$/WTe$_2$ heterostructure reference initialized in plane. The GROGU calculations used a $60\times60\times1$ reciprocal-space grid and a conservative energy-integration discretization of 300 points, used uniformly for all parameter sets. Exchange tensors were evaluated for Cr--Cr pairs up to the interaction range used in the Monte Carlo simulations, giving 114 pair interactions in the extracted parameter files. For the analysis shown in the main text, the isotropic interactions were shell averaged as a function of Cr--Cr distance, whereas the full bond-resolved tensors were retained for the VAMPIRE simulations.
In the torque evaluations, the Fermi level was not kept fixed externally; with the automatic electronic-temperature setting, GROGU redetermined the chemical potential from the Brillouin-zone integration for the corresponding rotated configuration.

 The complete bond-resolved exchange tensors, onsite anisotropy matrices and GROGU input files are provided in the Supplementary Information and in the accompanying data repository.

 \subsection*{VAMPIRE}

Finite-temperature magnetic properties were simulated with the VAMPIRE atomistic spin-dynamics package \cite{Evans2014_VAMPIRE}. The Cr magnetic degrees of freedom of the CrI$_3$ layer were represented by classical unit spin vectors placed at the Cr positions of the structural model and periodically replicated using custom VAMPIRE unit-cell files. The relativistic spin-Hamiltonian parameters extracted with GROGU were converted into the VAMPIRE input format. Pairwise interactions were included through the bond-resolved exchange tensors generated from the GROGU output, retaining the isotropic, symmetric anisotropic, and antisymmetric exchange contributions within the interaction cutoff used for the extracted parameter sets.

The onsite anisotropy obtained from GROGU was represented in the uniaxial form required by the VAMPIRE material input. For each parameter set, the principal anisotropy direction and magnitude were obtained from the corresponding onsite anisotropy tensor and used to define the local anisotropy axis and energy scale in the Monte Carlo simulations. This approximation affects only the representation of the onsite term in the VAMPIRE input; the mechanism of the substrate-induced reorientation was analyzed directly from the full GROGU tensors in Table~\ref{tab:GROGU_anisotropy}. The bond-resolved exchange tensors, including the two-ion relativistic anisotropic terms, were retained in the atomistic spin model.

Monte Carlo simulations were performed for finite periodically replicated Cr spin systems with lateral sizes of $30$, $50$, and $70$~nm in order to check the robustness of the transition temperature against simulation size. At each temperature, the system was equilibrated for $10,000$ Monte Carlo steps, followed by $5,000$ averaging steps. Temperature scans were performed with a step of $0.5$~K over a range covering the magnetic transition. Independent runs with different random seeds were used to verify that the extracted ordering temperatures were not seed-specific.

The magnetic order parameter was computed as
\[
m(T)=\frac{\langle |\mathbf M| \rangle_T}{M_0},
\qquad
\mathbf M=\frac{1}{N}\sum_{i=1}^N \mathbf e_i ,
\]
where $\langle\cdots\rangle_T$ denotes the Monte Carlo thermal average and $M_0$ is the low-temperature saturation value.

This choice is appropriate for comparing the isolated out-of-plane CrI$_3$ reference with the CrI$_3$/WTe$_2$ heterostructure, where the ordered moment lies in plane. The magnetic susceptibility was computed from magnetization fluctuations, and the ordering temperature was estimated from the peak position of the susceptibility curve. The simulation input files, converted VAMPIRE material and unit-cell files, finite-size checks, and susceptibility curves are provided in the accompanying data repository.

\subsection*{Workflow and reproducibility}

The complete workflow starts from a relaxed structural model, for which a self-consistent spin-orbit SIESTA calculation is performed while saving the Hamiltonian, overlap, density-matrix, and structural output files required for post-processing. GROGU then reads these SIESTA outputs, extracts the relativistic spin-Hamiltonian parameters using the LKAG-based mapping described above, and decomposes the interactions into isotropic exchange, symmetric anisotropic exchange, Dzyaloshinskii--Moriya vectors, and onsite anisotropy tensors. Importantly, GROGU automatically writes spin-dynamics-ready output files compatible with atomistic simulation codes such as VAMPIRE and UppASD, so that the extracted ab initio interactions can be used directly in finite-temperature simulations after choosing the desired simulation cell, temperature range, and Monte Carlo or spin-dynamics protocol. To make the workflow reproducible, the Supplementary Information and accompanying data repository contain the SIESTA input files, GROGU extraction inputs and outputs, converted VAMPIRE files and the data used to generate the results.

\section*{Data availability}

The data supporting the findings of this study are available in Zenodo under DOI: \url{https://doi.org/10.5281/zenodo.20645208}. The repository contains the SIESTA input and output files, pseudopotentials, GROGU spin-Hamiltonian extraction inputs and outputs, VAMPIRE Monte Carlo input and output files, and the numerical data used to generate the main manuscript figures. The dataset follows the complete SIESTA--GROGU--VAMPIRE workflow used in this work, from spin-orbit DFT calculations through relativistic spin-Hamiltonian extraction to finite-temperature atomistic Monte Carlo simulations.

\section*{Code availability}

The first-principles electronic-structure calculations were performed with the open-source SIESTA code \cite{Soler2002_SIESTA,Garcia2020_SIESTA}. Relativistic spin-Hamiltonian parameters were extracted using the publicly available GROGU package \cite{zerothi_grogupy}, with documentation available at \url{https://GROGU.readthedocs.io/en/stable/}. Atomistic Monte Carlo simulations were performed with VAMPIRE \cite{Evans2014_VAMPIRE}. The data-conversion files, and figure-generation data used in this work are included in the Zenodo dataset associated with this manuscript under DOI: \url{https://doi.org/10.5281/zenodo.20645208}.

\section*{Acknowledgements}

Supported by the DKOP-23 Doctoral Excellence Program of the Ministry for Culture and Innovation from the source of the National Research, Development and Innovation Fund. This work was supported by the National Research, Development and Innovation Office (NKFIH) in Hungary, through Grant No. FK-142985, K 146156 and Excellence 151372 and by the Hungarian Academy of Sciences LP2024-17 Lend\"{u}let ``Momentum'' grant. Z.T. acknowledges support from the János Bolyai Research Scholarship of the Hungarian Academy of Sciences. This project is supported by the TRILMAX Horizon Europe consortium (Grant No. 101159646) and Ministry of Culture and Innovation and the National Research, Development and Innovation Office under No. K142652 and ADVANCED 149745. We acknowledge the Digital Government Development and Project Management Ltd. for awarding us access to the Komondor HPC facility based in Hungary. This research has been funded by MCIN/AEI/10.13039/501100011033/ FEDER, UE via 
project PID2022-137078NB-100 and by Agencia SEKUENS (Asturias) under grant UONANO IDE/2024/000678 with the support of FEDER funds.

\section*{Author contributions}

Z.T. conceived the project, coordinated the work, performed the SIESTA DFT calculations, analyzed the electronic-structure results. D.T.P. developed and performed the GROGU spin-Hamiltonian extraction workflow and wrote the manuscript with input from all authors M.B.S. performed the VAMPIRE atomistic Monte Carlo simulations and analyzed the finite-temperature magnetic properties. L.O. contributed to the spin-model formulation, interpretation of the relativistic magnetic interactions, and overall theoretical strategy. A.G.-F. generated and validated the pseudopotentials used in the SIESTA calculations. L.S., V.I. and J.F. provided theoretical guidance on relativistic magnetic interactions and spin-Hamiltonian mapping. E.N.-M. and P.N.-I. contributed the experimental perspective and helped place the results in the broader context of two-dimensional magnetic heterostructures. All authors discussed the results and contributed to the final manuscript.

\section*{Competing interests}

The authors declare no competing interests.


\bibliography{cikkek} 

@article{WebsterYan2018_CrX3StrainMAE,
  author  = {Webster, Lucas and Yan, Jia-An},
  title   = {Strain-tunable magnetic anisotropy in monolayer {CrCl}$_3$, {CrBr}$_3$, and {CrI}$_3$},
  journal = {Physical Review B},
  volume  = {98},
  number  = {14},
  pages   = {144411},
  year    = {2018},
  doi     = {10.1103/PhysRevB.98.144411}
}

@article{Vishkayi2020_CrI3StrainElectricField,
  author  = {Vishkayi, Sahar Izadi and Torbatian, Zahra and Qaiumzadeh, Alireza and Asgari, Reza},
  title   = {Strain and electric-field control of spin-spin interactions in monolayer {CrI}$_3$},
  journal = {Physical Review Materials},
  volume  = {4},
  number  = {9},
  pages   = {094004},
  year    = {2020},
  doi     = {10.1103/PhysRevMaterials.4.094004}
}

@article{Wu2019_CrI3Strain,
  author  = {Wu, Zewen and Yu, Jin and Yuan, Shengjun},
  title   = {Strain-tunable magnetic and electronic properties of monolayer {CrI}$_3$},
  journal = {Physical Chemistry Chemical Physics},
  volume  = {21},
  number  = {15},
  pages   = {7750--7755},
  year    = {2019},
  doi     = {10.1039/C8CP07067A}
}

@article{Pizzochero2020_CrI3LatticeDeformations,
  author  = {Pizzochero, Michele and Yazyev, Oleg V.},
  title   = {Inducing Magnetic Phase Transitions in Monolayer {CrI}$_3$ via Lattice Deformations},
  journal = {The Journal of Physical Chemistry C},
  volume  = {124},
  number  = {13},
  pages   = {7585--7590},
  year    = {2020},
  doi     = {10.1021/acs.jpcc.0c01873}
}

@article{WangSanyal2021_CrI3Stacking,
  author  = {Wang, Duo and Sanyal, Biplab},
  title   = {Systematic Study of Monolayer to Trilayer {CrI}$_3$: Stacking Sequence Dependence of Electronic Structure and Magnetism},
  journal = {The Journal of Physical Chemistry C},
  volume  = {125},
  number  = {33},
  pages   = {18467--18473},
  year    = {2021},
  doi     = {10.1021/acs.jpcc.1c04311}
}

@article{MartinezCarracedo2023_RelativisticMagneticInteractions,
  author  = {Mart{\'i}nez-Carracedo, Gabriel and Oroszl{\'a}ny, L{\'a}szl{\'o} and Garc{\'i}a-Fuente, Amador and Ny{\'a}ri, Bendeg{\'u}z and Udvardi, L{\'a}szl{\'o} and Szunyogh, L{\'a}szl{\'o} and Ferrer, Jaime},
  title   = {Relativistic magnetic interactions from nonorthogonal basis sets},
  journal = {Physical Review B},
  volume  = {108},
  number  = {21},
  pages   = {214418},
  year    = {2023},
  doi     = {10.1103/PhysRevB.108.214418}
}

@article{Liechtenstein1987_LKAGExchange,
  author  = {Liechtenstein, A. I. and Katsnelson, M. I. and Antropov, V. P. and Gubanov, V. A.},
  title   = {Local spin density functional approach to the theory of exchange interactions in ferromagnetic metals and alloys},
  journal = {Journal of Magnetism and Magnetic Materials},
  volume  = {67},
  number  = {1},
  pages   = {65--74},
  year    = {1987},
  doi     = {10.1016/0304-8853(87)90721-9}
}

@article{Soler2002_SIESTA,
  author  = {Soler, J. M. and Artacho, E. and Gale, J. D. and Garc{\'i}a, A. and Junquera, J. and Ordej{\'o}n, P. and S{\'a}nchez-Portal, D.},
  title   = {The {SIESTA} method for ab initio order-{N} materials simulation},
  journal = {Journal of Physics: Condensed Matter},
  volume  = {14},
  number  = {11},
  pages   = {2745--2779},
  year    = {2002},
  doi     = {10.1088/0953-8984/14/11/302}
}

@article{Garcia2020_SIESTA,
  author  = {Garc{\'i}a, A. and Papior, N. and Akhtar, A. and Artacho, E. and Blum, V. and Bosoni, E. and Brandimarte, P. and Brandbyge, M. and Cerd{\'a}, J. I. and Corsetti, F. and Cuadrado, R. and Dikan, V. and Ferrer, J. and Gale, J. and Garc{\'i}a-Fern{\'a}ndez, P. and Garc{\'i}a-Su{\'a}rez, V. M. and Garc{\'i}a, S. and Huhs, G. and Illera, S. and Koryt{\'a}r, R. and Koval, P. and Lebedeva, I. and Lin, L. and L{\'o}pez-Tarifa, P. and Mayo, S. G. and Mohr, S. and Ordej{\'o}n, P. and Postnikov, A. and Pouillon, Y. and Pruneda, M. and Robles, R. and S{\'a}nchez-Portal, D. and Soler, J. M. and Ullah, R. and Yu, V. Wen-zhe and Junquera, J.},
  title   = {{Siesta}: Recent developments and applications},
  journal = {The Journal of Chemical Physics},
  volume  = {152},
  number  = {20},
  pages   = {204108},
  year    = {2020},
  doi     = {10.1063/5.0005077}
}

@article{Perdew1996_PBE,
  author  = {Perdew, John P. and Burke, Kieron and Ernzerhof, Matthias},
  title   = {Generalized Gradient Approximation Made Simple},
  journal = {Physical Review Letters},
  volume  = {77},
  number  = {18},
  pages   = {3865--3868},
  year    = {1996},
  doi     = {10.1103/PhysRevLett.77.3865}
}

@article{Evans2014_VAMPIRE,
  author  = {Evans, R. F. L. and Fan, W. J. and Chureemart, P. and Ostler, T. A. and Ellis, M. O. A. and Chantrell, R. W.},
  title   = {Atomistic spin model simulations of magnetic nanomaterials},
  journal = {Journal of Physics: Condensed Matter},
  volume  = {26},
  number  = {10},
  pages   = {103202},
  year    = {2014},
  doi     = {10.1088/0953-8984/26/10/103202}
}

@software{zerothi_grogupy,
  author  = {Pozs{\'a}r, D{\'a}niel Tibor and Mart{\'i}nez-Carracedo, Gabriel and Garc{\'i}a-Fuente, Amador and Udvardi, L{\'a}szl{\'o} and Szunyogh, L{\'a}szl{\'o} and Ferrer, Jaime and Oroszl{\'a}ny, L{\'a}szl{\'o}},
  title   = {{grogupy}: v0.4.0},
  year    = {2025},
  doi     = {10.5281/zenodo.15449541},
  url     = {https://doi.org/10.5281/zenodo.15449541}
}

@article{Gong2017_CrGeTe3Intrinsic2DFerromagnetism,
  author  = {Gong, Cheng and Li, Lin and Li, Zhenglu and Ji, Huiwen and Stern, Alex and Xia, Yang and Cao, Ting and Bao, Wei and Wang, Chenzhe and Wang, Yuan and Qiu, Zhi Q. and Cava, R. J. and Louie, Steven G. and Xia, Jing and Zhang, Xiang},
  title   = {Discovery of intrinsic ferromagnetism in two-dimensional van der Waals crystals},
  journal = {Nature},
  volume  = {546},
  pages   = {265--269},
  year    = {2017},
  doi     = {10.1038/nature22060}
}

@article{Huang2017_LayerDependentCrI3Ferromagnetism,
  author  = {Huang, Bevin and Clark, Genevieve and Navarro-Moratalla, Efr{\'e}n and Klein, David R. and Cheng, Ran and Seyler, Kyle L. and Zhong, Ding and Schmidgall, Emma and McGuire, Michael A. and Cobden, David H. and Yao, Wang and Xiao, Di and Jarillo-Herrero, Pablo and Xu, Xiaodong},
  title   = {Layer-dependent ferromagnetism in a van der Waals crystal down to the monolayer limit},
  journal = {Nature},
  volume  = {546},
  pages   = {270--273},
  year    = {2017},
  doi     = {10.1038/nature22391}
}

@article{Burch2018_Magnetism2DvdWMaterials,
  author  = {Burch, Kenneth S. and Mandrus, David and Park, Je-Geun},
  title   = {Magnetism in two-dimensional van der Waals materials},
  journal = {Nature},
  volume  = {563},
  pages   = {47--52},
  year    = {2018},
  doi     = {10.1038/s41586-018-0631-z}
}

@article{Gibertini2019_Magnetic2DMaterialsHeterostructures,
  author  = {Gibertini, Marco and Koperski, Maciej and Morpurgo, Alberto F. and Novoselov, Kostya S.},
  title   = {Magnetic 2D materials and heterostructures},
  journal = {Nature Nanotechnology},
  volume  = {14},
  pages   = {408--419},
  year    = {2019},
  doi     = {10.1038/s41565-019-0438-6}
}

@article{Lado2017_CrI3MagneticAnisotropyOrigin,
  author  = {Lado, J. L. and Fern{\'a}ndez-Rossier, J.},
  title   = {On the origin of magnetic anisotropy in two dimensional {CrI}$_3$},
  journal = {2D Materials},
  volume  = {4},
  number  = {3},
  pages   = {035002},
  year    = {2017},
  doi     = {10.1088/2053-1583/aa75ed}
}

@article{Kashin2020_OrbitallyResolvedCrI3Ferromagnetism,
  author  = {Kashin, I. V. and Mazurenko, V. V. and Katsnelson, M. I. and Rudenko, A. N.},
  title   = {Orbitally-resolved ferromagnetism of monolayer {CrI}$_3$},
  journal = {2D Materials},
  volume  = {7},
  number  = {2},
  pages   = {025036},
  year    = {2020},
  doi     = {10.1088/2053-1583/ab72d8}
}

@article{JaeschkeUbiergo2021_CrI3MagnetismTheory,
  author  = {Jaeschke-Ubiergo, Rodrigo and Su{\'a}rez Morell, Eric and N{\'u}{\~n}ez, Alvaro S.},
  title   = {Theory of magnetism in the van der Waals magnet {CrI}$_3$},
  journal = {Physical Review B},
  volume  = {103},
  number  = {17},
  pages   = {174410},
  year    = {2021},
  doi     = {10.1103/PhysRevB.103.174410}
}

@article{Bacaksiz2021_CrI3CrBr3SOCMagneticProperties,
  author  = {Bacaksiz, Cihan and {\v S}abani, Denis and Menezes, Ra{\'i} M. and Milo{\v s}evi{\'c}, Milorad V.},
  title   = {Distinctive magnetic properties of {CrI}$_3$ and {CrBr}$_3$ monolayers caused by spin-orbit coupling},
  journal = {Physical Review B},
  volume  = {103},
  number  = {12},
  pages   = {125418},
  year    = {2021},
  doi     = {10.1103/PhysRevB.103.125418}
}

@article{Sabani2025_BeyondOrbitallyResolvedExchangeCrI3NiI2,
  author  = {{\v S}abani, Denis and Bacaks{\i}z, Cihan and Milo{\v s}evi{\'c}, Milorad V.},
  title   = {Beyond Orbitally Resolved Magnetic Exchange in {CrI}$_3$ and {NiI}$_2$},
  journal = {Physical Review Letters},
  volume  = {135},
  number  = {3},
  pages   = {036704},
  year    = {2025},
  doi     = {10.1103/tlq2-m6zk}
}

@article{Tiwari2021_CriticalBehaviorCrI3CrBr3CrGeTe3FeCl2,
  author  = {Tiwari, Sabyasachi and Van de Put, Maarten L. and Sor{\'e}e, Bart and Vandenberghe, William G.},
  title   = {Critical behavior of the ferromagnets {CrI}$_3$, {CrBr}$_3$, and {CrGeTe}$_3$ and the antiferromagnet {FeCl}$_2$: A detailed first-principles study},
  journal = {Physical Review B},
  volume  = {103},
  number  = {1},
  pages   = {014432},
  year    = {2021},
  doi     = {10.1103/PhysRevB.103.014432}
}

@article{Pavizhakumari2025_BeyondRPACurieTemperatureCrI3,
  author  = {Pavizhakumari, Varun Rajeev and Skovhus, Thorbj{\o}rn and Olsen, Thomas},
  title   = {Beyond the random phase approximation for calculating Curie temperatures in ferromagnets: application to {Fe}, {Ni}, {Co} and monolayer {CrI}$_3$},
  journal = {Journal of Physics: Condensed Matter},
  volume  = {37},
  number  = {11},
  pages   = {115806},
  year    = {2025},
  doi     = {10.1088/1361-648X/ada65c}
}

@article{Dolui2020_ProximitySpinOrbitTorqueCrI3,
  author  = {Dolui, Kapildeb and Petrovi{\'c}, Marko D. and Zollner, Klaus and Plech{\'a}{\v c}, Petr and Fabian, Jaroslav and Nikoli{\'c}, Branislav K.},
  title   = {Proximity Spin--Orbit Torque on a Two-Dimensional Magnet within van der Waals Heterostructure: Current-Driven Antiferromagnet-to-Ferromagnet Reversible Nonequilibrium Phase Transition in Bilayer {CrI}$_3$},
  journal = {Nano Letters},
  volume  = {20},
  number  = {4},
  pages   = {2288--2295},
  year    = {2020},
  doi     = {10.1021/acs.nanolett.9b04556}
}

@article{Zollner2019_ProximityExchangeMoSe2WSe2CrI3,
  author  = {Zollner, Klaus and Faria Junior, Paulo E. and Fabian, Jaroslav},
  title   = {Proximity exchange effects in {MoSe}$_2$ and {WSe}$_2$ heterostructures with {CrI}$_3$: Twist angle, layer, and gate dependence},
  journal = {Physical Review B},
  volume  = {100},
  number  = {8},
  pages   = {085128},
  year    = {2019},
  doi     = {10.1103/PhysRevB.100.085128}
}

@article{Zollner2023_ValleySplittingTwistingGatingCrI3,
  author  = {Zollner, Klaus and Faria Junior, Paulo E. and Fabian, Jaroslav},
  title   = {Strong manipulation of the valley splitting upon twisting and gating in {MoSe}$_2$/{CrI}$_3$ and {WSe}$_2$/{CrI}$_3$ van der Waals heterostructures},
  journal = {Physical Review B},
  volume  = {107},
  number  = {3},
  pages   = {035112},
  year    = {2023},
  doi     = {10.1103/PhysRevB.107.035112}
}

@article{Staros2024_WTe2CrI3TopologicalSpinFilter,
  author  = {Staros, Daniel and Rubenstein, Brenda and Ganesh, Panchapakesan},
  title   = {A first-principles study of bilayer 1T'-{WTe}$_2$/{CrI}$_3$: a candidate topological spin filter},
  journal = {npj Spintronics},
  volume  = {2},
  pages   = {4},
  year    = {2024},
  doi     = {10.1038/s44306-023-00007-y}
}

@article{Zhao2020_MagneticProximityWTe2HelicalEdge,
  author  = {Zhao, Wenjin and Fei, Zaiyao and Song, Tiancheng and Choi, Han Kyou and Palomaki, Tauno and Sun, Bosong and Malinowski, Paul and McGuire, Michael A. and Chu, Jiun-Haw and Xu, Xiaodong and Cobden, David H.},
  title   = {Magnetic proximity and nonreciprocal current switching in a monolayer {WTe}$_2$ helical edge},
  journal = {Nature Materials},
  volume  = {19},
  number  = {5},
  pages   = {503--507},
  year    = {2020},
  doi     = {10.1038/s41563-020-0620-0}
}

@article{Ma2016_HexagonalTMDQuantumSpinHall,
  author  = {Ma, Yandong and Kou, Liangzhi and Li, Xiao and Dai, Ying and Heine, Thomas},
  title   = {Two-dimensional transition metal dichalcogenides with a hexagonal lattice: Room-temperature quantum spin Hall insulators},
  journal = {Physical Review B},
  volume  = {93},
  number  = {3},
  pages   = {035442},
  year    = {2016},
  doi     = {10.1103/PhysRevB.93.035442}
}

@article{Ando2024_SelectiveFabricationMonolayerWTe2,
  author  = {Ando, Ryuichi and Sugawara, Katsuaki and Kawakami, Tappei and Takahashi, Takashi and Sato, Takafumi},
  title   = {Selective Fabrication of Monolayer 1H- and 1T'-{WTe}$_2$},
  journal = {Journal of the Physical Society of Japan},
  volume  = {93},
  number  = {8},
  pages   = {085002},
  year    = {2024},
  doi     = {10.7566/JPSJ.93.085002},
  eprint  = {2501.17527},
  archivePrefix = {arXiv}
}

@article{Zhou2026_WTe2Fe3GaTe2SOT,
  author  = {Zhou, Sicheng and Yang, Jiakai and Hu, Guojing and Chen, Zanhong and Shi, K. and Tong, B. and Feng, J. and Dou, Z. and Lyu, Z. and Song, X. and Shen, J. and Yang, H. and Ke, H. and Li, P. and Liu, G. and Qu, F. and Lu, L.},
  title   = {Field-free and efficient nanosecond spin--orbit torque switching in optimized {WTe}$_2$/{Fe}$_3${GaTe}$_2$ heterostructures},
  journal = {Applied Physics Letters},
  volume  = {128},
  number  = {19},
  pages   = {192403},
  year    = {2026},
  doi     = {10.1063/5.0327470}
}

@article{Herling2026_CGTWTe2Ferromagnetism,
  author  = {Herling, Franz and Torres-Sala, Mireia and Esteras, Dorye L. and Evason, Charlotte and Aoki, Motomi and Rosado, Marcos and Gupta, Kapil and Mundet, Bernat and Xu, Kai and Reparaz, J. Sebasti{\'a}n and Watanabe, Kenji and Taniguchi, Takashi and Dimitrov, Dimitre and Marinova, Vera and Verzhbitskiy, Ivan A. and Eda, Goki and Garcia, Jos{\'e} H. and Roche, Stephan and Sierra, Juan F. and Valenzuela, Sergio O.},
  title   = {Strain-mediated lattice reconstruction enhances ferromagnetism in {Cr}$_2${Ge}$_2${Te}$_6$/{WTe}$_2$ van der {W}aals heterobilayers},
  journal = {Nano Letters},
  volume  = {26},
  number  = {16},
  pages   = {5434--5442},
  year    = {2026},
  doi     = {10.1021/acs.nanolett.6c00201}
}

@article{Zhang2026_vdWSOTImaging,
  author  = {Zhang, Xi and Zhou, Jingcheng and Hu, Chaowei and Cai, Peiyu and Deng, Kuangyin and Wang, Chuangtang and Agarwal, Nishkarsh and Jin, Hanshang and Al-Matouq, Faris A. and Xu, Stelo and Trivedi, Roshan S. and Wang, Hailong},
  title   = {Imaging of a van der Waals spin-orbit torque system using spin ensembles in hBN},
  journal = {Nature Communications},
  year    = {2026},
  doi     = {10.1038/s41467-026-74178-7}
}

@misc{Ning2024_WTe2Fe3GeTe2SpinTransfer,
  author        = {Ning, H. L. and Dong, M. Q. and Zhang, X. and Huang, J. S. and Liu, B. and Guo, Zhi-Xin and Dong, Y.},
  title         = {Efficient Spin Transfer in {WTe}$_2$/{Fe}$_3${GeTe}$_2$ van der {W}aals Heterostructure Enabled by Direct Interlayer p-Orbital Hybridization},
  year          = {2024},
  eprint        = {2412.02966},
  archivePrefix = {arXiv},
  primaryClass  = {cond-mat.mtrl-sci},
  doi           = {10.48550/arXiv.2412.02966}
}

@article{Li2022_ProximityMagnetizedQSHI,
  author  = {Li, Junxue and Rashetnia, Mina and Lohmann, Mark and Koo, Jahyun and Xu, Youming and Zhang, Xiao and Watanabe, Kenji and Taniguchi, Takashi and Jia, Shuang and Chen, Xi and Yan, Binghai and Cui, Yong-Tao and Shi, Jing},
  title   = {Proximity-magnetized quantum spin Hall insulator: monolayer 1T$^\prime$-{WTe}$_2$/{Cr}$_2${Ge}$_2${Te}$_6$},
  journal = {Nature Communications},
  volume  = {13},
  pages   = {5134},
  year    = {2022},
  doi     = {10.1038/s41467-022-32808-w}
}

@article{HenriquezGuerra2025_CrSBrStrainMRMAE,
  author  = {Henr{\'i}quez-Guerra, E. and others},
  title   = {Strain Engineering of Magnetoresistance and Magnetic Anisotropy in {CrSBr}},
  journal = {Advanced Materials},
  pages   = {e2506695},
  year    = {2025},
  doi     = {10.1002/adma.202506695}
}

@article{Cheng2022_ElectricFieldTunableInterfacialFerromagnetism,
  author  = {Cheng, Guanghui and Rahman, Mohammad Mushfiqur and He, Zhiping and Allcca, Andres Llacsahuanga and Rustagi, Avinash and Stampe, Kirstine A. and Zhu, Yanglin and Yan, Shaohua and Tian, Shangjie and Mao, Zhiqiang and Lei, Hechang and Watanabe, Kenji and Taniguchi, Takashi and Upadhyaya, Pramey and Chen, Yong P.},
  title   = {Emergence of electric-field-tunable interfacial ferromagnetism in 2D antiferromagnet heterostructures},
  journal = {Nature Communications},
  volume  = {13},
  pages   = {7348},
  year    = {2022},
  doi     = {10.1038/s41467-022-34812-6}
}

@article{Eom2023_VoltageControlMagnetism,
  author  = {Eom, Jaeun and Lee, I. H. and Kee, J. Y. and Cho, Minhyun and Seo, Jeongdae and Suh, Hoyoung and Choi, Hyung-Jin and Sim, Yumin and Chen, Shuzhang and Chang, Hye Jung and others},
  title   = {Voltage control of magnetism in {Fe}$_{3-x}${GeTe}$_2$/{In}$_2${Se}$_3$ van der Waals ferromagnetic/ferroelectric heterostructures},
  journal = {Nature Communications},
  volume  = {14},
  pages   = {5605},
  year    = {2023},
  doi     = {10.1038/s41467-023-41382-8}
}

@article{Udvardi2003_RelativisticSpinWaves,
  author  = {Udvardi, L. and Szunyogh, L. and Palot{\'a}s, K. and Weinberger, P.},
  title   = {First-principles relativistic study of spin waves in thin magnetic films},
  journal = {Physical Review B},
  volume  = {68},
  pages   = {104436},
  year    = {2003},
  doi     = {10.1103/PhysRevB.68.104436}
}

@article{Krindges2026_ElectronicCorrelationsCrI3,
  author  = {Krindges, Arthur and
             Vaz de Morais Junior, Carlos Alberto and
             Piotrowski, Maur{\'i}cio Jeomar},
  title   = {Role of Electronic Correlations on Exchange Interactions and
             Curie Temperature in Monolayer {CrI}$_3$},
  journal = {ACS Omega},
  year    = {2026},
  volume  = {11},
  number  = {22},
  doi     = {10.1021/acsomega.6c02634},
  url     = {https://doi.org/10.1021/acsomega.6c02634},
  publisher = {American Chemical Society}
}

\end{document}